\documentclass[journal]{IEEEtran}
\pdfoutput=1
\usepackage{color}
\usepackage{booktabs}
\usepackage{arydshln}
\usepackage{multirow}
\usepackage{ulem}
\usepackage[pdftex]{graphicx}
\graphicspath{{../pdf/}{../jpeg/}}
\DeclareGraphicsExtensions{.pdf,.jpeg,.png}
\usepackage[caption=false,font=footnotesize]{subfig}
\begin{document}

\title{\textcolor{black}{Sequential Weakly Labeled Multi-Activity Localization and Recognition on Wearable Sensors using Recurrent Attention Networks}}

\author{Kun~Wang,
		Jun~He,~\IEEEmembership{Member,~IEEE}
        and~Lei~Zhang%
\thanks{The work is supported in part by the National Natural Science Foundation of China under Grant 61203237 and the Natural Science Foundation of Jiangsu Province under grant BK20191371.\textit{ (Corresponding author: Lei Zhang.)}}%
\thanks{Kun Wang and Lei Zhang are with School of Electrical and Automation Engineering, Nanjing Normal University, Nanjing, 210023, China (e-mail: iskenn7@gmail.com, leizhang@njnu.edu.cn). }%
\thanks{Jun He is with the School of Electronic and Information Engineering, Nanjing University of Information Science and Technology, Nanjing, 210044, China (e-mail: jhe@nuist.edu.cn). }%
}

\maketitle

\begin{abstract}
With the popularity and development of the wearable devices such as smartphones, human activity recognition (HAR) based on sensors has become as a key research area in human computer interaction and ubiquitous computing. The emergence of deep learning leads to a recent shift in the research of HAR, which requires massive strictly labeled data. In comparison with video data, activity data recorded from accelerometer or gyroscope is often more difficult to interpret and segment. Recently, several attention mechanisms are proposed to handle the weakly labeled human activity data, which do not require accurate data annotation. However, these attention-based models can only handle the weakly labeled dataset whose sample includes one target activity, as a result it limits efficiency and practicality. In the paper, we propose a recurrent attention networks (RAN) to handle sequential weakly labeled multi-activity recognition and location tasks. The model can repeatedly perform steps of attention on multiple activities of one sample and each step is corresponding to the current focused activity. The effectiveness of the RAN model is validated on a collected sequential weakly labeled multi-activity dataset and the other two public datasets. The experiment results show that our RAN model can simultaneously infer multi-activity types from the coarse-grained sequential weak labels and determine specific locations of every target activity with only knowledge of which types of activities contained in the long sequence. It will greatly reduce the burden of manual labeling. The code of our work is available at https://github.com/KennCoder7/RAN.
\end{abstract}

\begin{IEEEkeywords}
human activity recognition, weakly labeled data, wearable sensors, recurrent attention networks
\end{IEEEkeywords}

\IEEEpeerreviewmaketitle

\section{Introduction}
\IEEEPARstart{H}{uman} activity is unique, as the information inferred from raw activity data has been proved to be very critical in human activity recognition (HAR) \cite{bulling2014tutorial}, health support \cite{magherini2013using}, and smart homes \cite{rashidi2009keeping} to name a few. With the popularity and development of the wearable devices such as smartphones, human activity can be captured using a variety of motion sensors such as accelerometer and gyroscope worn on various parts of the body, which provides convenient interface between humans and machines \cite{ foerster1999detection, mannini2010machine, pei2013human}. HAR can perform automatic detection of various physical activities performed by people in their daily lives. The traditional machine learning approaches such as Support Vector Machine and Hidden Markov Model, based on hand-crafted features \cite{bao2004activity, huynh2007scalable, kwapisz2011activity, anguita2012human}, have been extensively used in the HAR field. 
\\
\indent
Due to the emergence of deep learning, there has been a recent shift in the use of machine learning techniques. Deep learning methods can learn the features automatically from the data which avoids the problem of the hand-crafted features in shallow learning field. Deep learning approaches such as Convolutional Neural Networks (CNN) \cite{zeng2014convolutional, chen2015deep, jiang2015human, yang2015deep} and Recurrent Neural Networks (RNN) \cite{ordonez2016deep, kumar2018multimodal, kim2018deepgesture} have proven to be more effective than the shallow learning techniques in discovering, learning, and inferring complex activity from data. These emerging methods, which are essentially inside the range of supervised learning, have achieved better performance in HAR. But there are some remaining challenges need to be addressed, the main one of which is how to build a well-labeled HAR dataset with ground truth annotation \cite{cruciani2018automatic}.
\begin{figure}[t]
	\centering
	\includegraphics[width=3.5in]{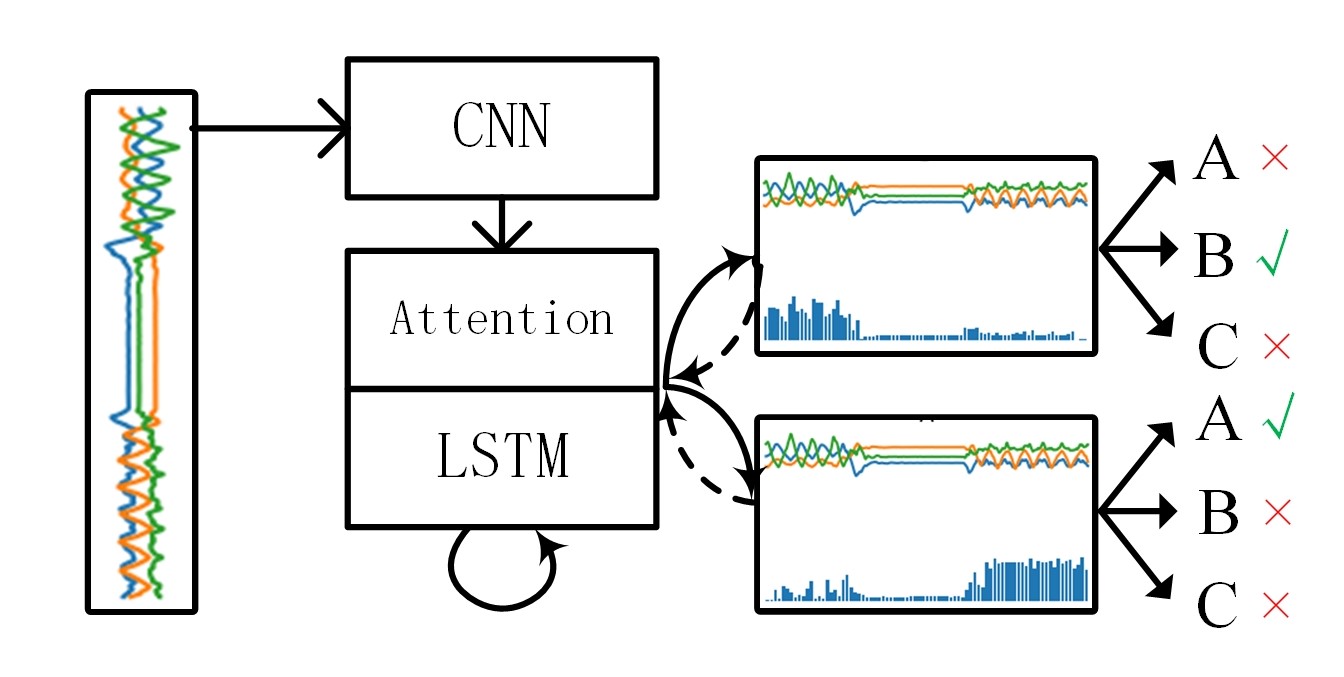}
	\caption{The overall architecture of our model, which can produce an attention map of the focused activity at every step.}
	\label{Fig. 1}
\end{figure}
\\
\indent
\textcolor{black}{ Intuitively, it is much easier for human beings who are recording sensor data to identify whether an interesting activity takes place in a long sequence. If we can infer activity types from the coarse-grained labels and determine specific locations of every labeled target activity with only knowledge of which types of activities contained in the long sequence, it will greatly reduce the burden of manual labeling. Therefore, it deserves further research whether we can directly recognize and locate one or multiple target activities from coarsely labeled sensor data. We tackle the above challenges from a novel aspect, i.e., attention, which recently has been studied extensively in various research areas such as computer vision and natural language processing (NLP).} In our previous work \cite{wang2019attention}, an end-to-end-trainable attention module was embedded into CNN architecture to identify interesting activity from weakly labeled dataset for HAR, saying that each sample in the dataset is composed of the interesting activity and the background activity. Without need of strict annotation, the attention-based CNN can greatly facilitate the process of sensor data collection. However, the attention-based CNN, can only deal with the weakly labeled samples that include one target activity, as a result it limits efficiency and practicality. Therefore, the new challenge is that whether one can simultaneously recognize and locate multiple target activities from one sample of the weakly labeled HAR dataset.
\\\indent
To tackle this challenge, one feasible proposal is that enabling the mechanism of attention to become “recurrent”. Recently, Xu et al. \cite{xu2015show} proposed a model to handle the task of automatically generating captions for an image, which can visualize “where” and “what” the attention focused on. The model can repeatedly perform steps of attention on multiple objects of an image and each step is corresponding to the current focused object. The model can generate the weighted feature of the current focused object according to its previous observations.
\\
\indent
In this paper, edified by this notion\cite{xu2015show}, we propose a recurrent attention networks (RAN) to recognize and locate multi-activity types in sequential weakly labeled sensor samples. \textcolor{black}{To the best of our knowledge, attention has seldom been combined with RNN in HAR scenario. This is the first paper to leverage RAN to deal with sequential weakly labeled sensor data.} The RAN model that consists of CNN, Long Short-Term Memory (LSTM) and attention module is shown in Fig. 1. For sequential activity recognition tasks, processing one sample usually consists of $T$ steps. At each step $t$, the model needs to produce an attention map of the current activity and its corresponding attention feature. Then the trained feature must change accordingly to represent different activities, therefore the recurrent structure can be naturally exploited to provide the conditional information for the variable feature.  
\textcolor{black}{\\\indent Our main contribution is three-fold. Firstly, we for the first time propose an efficient RAN model in HAR scenarios, which can simultaneously infer one or multiple activity kinds from the coarse-grained labels and determine specific locations of every target activity with only knowledge of what kinds of activities contained in the long sequence. Secondly, we perform extensive experiments to verify that the attention method achieves comparable performance with standard CNN, across the public benchmark UniMiB-SHAR and OPPORTUNITY datasets, as well as the sequential weakly labeled HAR dataset. Thirdly, visualizing analysis of attention weights along temporal dimension is provided to improve the comprehensibility of the sensor data annotation. The proposed method can effectively aid to the collection of “ground truth labeled” training data.}
\\
\indent
The remainder of this paper is structured as follows. An overview of related works appears in Section \uppercase\expandafter{\romannumeral2}. In Section \uppercase\expandafter{\romannumeral3}, we describe the SWLM dataset collected for this research and two public datasets. Section \uppercase\expandafter{\romannumeral4} propose our RAN model. The experimental results and discussion are then presented in Section \uppercase\expandafter{\romannumeral5}. Finally, the paper is concluded in Section \uppercase\expandafter{\romannumeral6}.

\section{Related Works}
\indent HAR, has emerged as a key research area in human computer interaction (HCI) and ubiquitous computing. HAR can be seen as a typical pattern recognition problem, which has made tremendous progress by adopting shallow learning algorithms. In \cite{bao2004activity}, Bao et al. found that accelerometer sensor data is suitable for activities recognition. Four types of features (mean, energy, frequency and domain entropy)
were extracted manually from accelerometer data and activity recognition on these features was performed using decision table, instance-based learning  (IBL or nearest neighbor), C4.5 decision tree, and Nave Bayes classifiers \cite{witten2016data}. Kwapisz et al. \cite{kwapisz2011activity} also used the accelerometer sensor of mobile devices to extract features, and six different hand-crafted features were generated and then fed into the classifiers
such as decision trees (J48), multi-layer perceptions (MLP), and logistic regression. However, the features of these shallow algorithms are usually extracted via a hand-crafted way, which heavily relies on domain knowledge or human experience and has low performance in distinguishing similar activities such as walking upstairs and walking downstairs \cite{ronao2015deep}. Besides, choosing suitable features and extracting features from sensor data manually are both difficult and laborious.
\\
\indent The emergence of deep learning tends to overcome above drawbacks, and the features can be learned automatically through convolutional networks instead of being manually designed \cite{lecun2015deep}. For instance, Chen and Xue \cite{chen2015deep} fed raw signal into a sophisticated CNN, which had an architecture composed of three convolutional layers and three max-pooling layers. Furthermore, Jiang and Yin \cite{jiang2015human} converted the raw sensor signal into 2D signal image by utilizing a specific permutation technique and discrete cosine transformation (DCT), then fed the 2D signal image into a two layer 2D CNN to classify the signal image equaling to its desired activity recognition. Ordonez et al. \cite{ordonez2016deep} proposed the DeepConvLSTM model comprised of CNN and LSTM recurrent units, which outperforms CNN. \textcolor{black}{Kumar et al. \cite{kumar2018multimodal} proposed an approach to perform multimodal gait recognition by fusing motion sensor and video data. The signals of the motion sensors are modeled using a LSTM network and corresponding video recordings are processed using a three-dimensional CNN. Finally, Gray wolf optimizer was used to optimize the parameters during fusion. Kim et al. \cite{kim2018deepgesture} proposed the deepGesture model to perform arm gesture recognition based on motion sensors. They used a CNN with four convolutional layers to learn features in raw sensor data, and then the features extracted by the CNN are used as an input of gated recurrent unit (GRU) to capture long-term dependency and model motion sensor signals. Here GRU can be seen as a lightweight LSTM structure.  Teng et al. \cite{teng2020layer} proposed a layer-wise CNN with local loss for the use of HAR task which shows that local loss works better than global loss for tested baseline architectures.  A state-of-the-art survey \cite{wang2019deep} on HAR based on wearable sensors depicted details about the performance of current deep learning models. } However, these methods that belong to supervised learning \cite{shoaib2015survey} requires massive data with perfect ground-truth to train models.
\\\indent As the annotator has to skim through the raw sensor data and manually label all activity instances, ground truth annotation is an expensive and tedious task. In comparison with other sensors, such as cameras, activity data recorded from an accelerometer or gyroscope is also often more difficult to interpret and segment. Strictly labeling samples of sensor data needs much more manpower and computing resources. Recently, some semi-supervised and weakly supervised learning approaches were proposed to improve the efficiency of the ground truth annotation tasks in HAR. Zeng et al. \cite{zeng2017semi} presented the semi-supervised methods based on CNN that can learn from both labeled and unlabeled data. Recent researches in computer vision \cite{jetley2018learn}, machine translation \cite{luong2015effective}, speech recognition \cite{chorowski2015attention}, and image caption \cite{xu2015show} have witnessed the success of attention mechanism. For example, in computer vision, attention does not need to focus on the whole image, but only on the salient areas of the image. The attention idea can be exploited to handle weakly labeled HAR data, which does not require the strict data annotation. Inspired by the notion, He et al. \cite{he2018weakly} proposed a weakly supervised model based on
recurrent attention learning, and this method can deal with
weakly labeled HAR data by utilizing an agent to adaptively select
from sequence of locations and then extract desired information. Besides,
in the previous work \cite{wang2019attention}, we proposed a soft attention CNN model via measuring the compatibility between local features and global features , which can amplify the salient activity information and suppress the irrelevant confusing information. Nevertheless, the above attention methods have the limitation that can only handle the weakly labeled samples whose segments include one labeled activity. 
\section{Dataset}
\subsection{UniMiB-SHAR Dataset}
\indent 
We utilize the public dataset to validate that our model can deal with traditional activity recognition tasks. The UniMiB-SHAR dataset consists of 17 daily activities aggregated from 30 volunteers. The data is recorded from a Samsung smartphone, which collects 3-axial linear acceleration at a constant rate of 50Hz. We use the method mentioned in \cite{li2018comparison} to preprocess this dataset. 30 volunteers’ data is divide into two parts where 20 subjects are for training and 10 for test. A fixed length window of 151 is used to segment the data.
\subsection{Sequential Weakly Labeled Multi-Activity Dataset}
\indent
\textcolor{black}{We collect the sequential weakly labeled multi-activity (SWLM) dataset to perform the sequential weakly labeled HAR tasks.} The dataset includes three types of activities: “chest stretch”, “arm lateral stretch” and “arm vertical stretch”, and we use “A”, “B” and “C” to denote the three activities. The sensors data is collected from 3-axis accelerometer of iPhone tied to 10 subjects' right wrist as shown in Fig. 2(a). The subjects do the above three actions in the order “A-B-C” and “C-B-A”. Each activity lasts five seconds (about five times), and there is a time gap between two types of activities. The smartphone has a sampling rate of 50Hz. As can be seen in Fig. 2(b), the whole process of collection is supported by a mobile application named HASC Logger \cite{kawaguchi2011hasc} which records the data from accelerometer and then uploads the data to computer terminal.
\\ \indent We divide raw data by distinguishing different subjects, and seven participants’ data is used for training and the rest three subjects’ data for test. Then we use a fixed length sliding window of 650 to segment the data. The whole process is illustrated in Fig. 3. Finally, the dataset consists of nine different types of samples: “A”, “B”, “C”, “A-B”, “B-C”, “C-A”, “C-B”, “B-A” and “A-C”. \textcolor{black}{Here, each sample should correspond to its sequential weak labels. For example, the sample "A-B" is assigned with the sequential label ("start, "A", "B", "end"). }
\begin{figure}[t]
	\centering
	\subfloat[]{\includegraphics[width=1in]{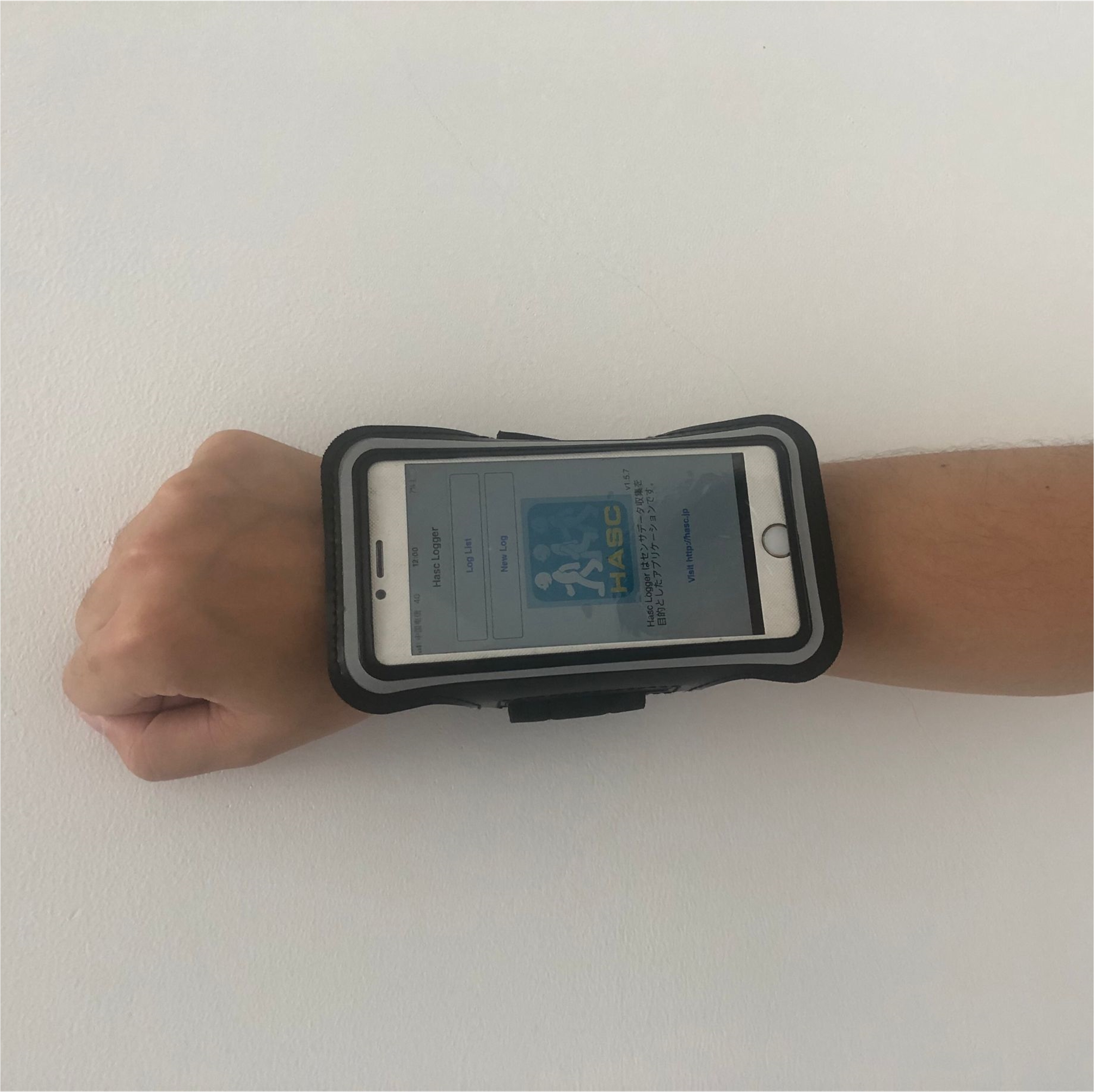}
		\label{Fig. 2(a)}}
	\subfloat[]{\includegraphics[width=2in]{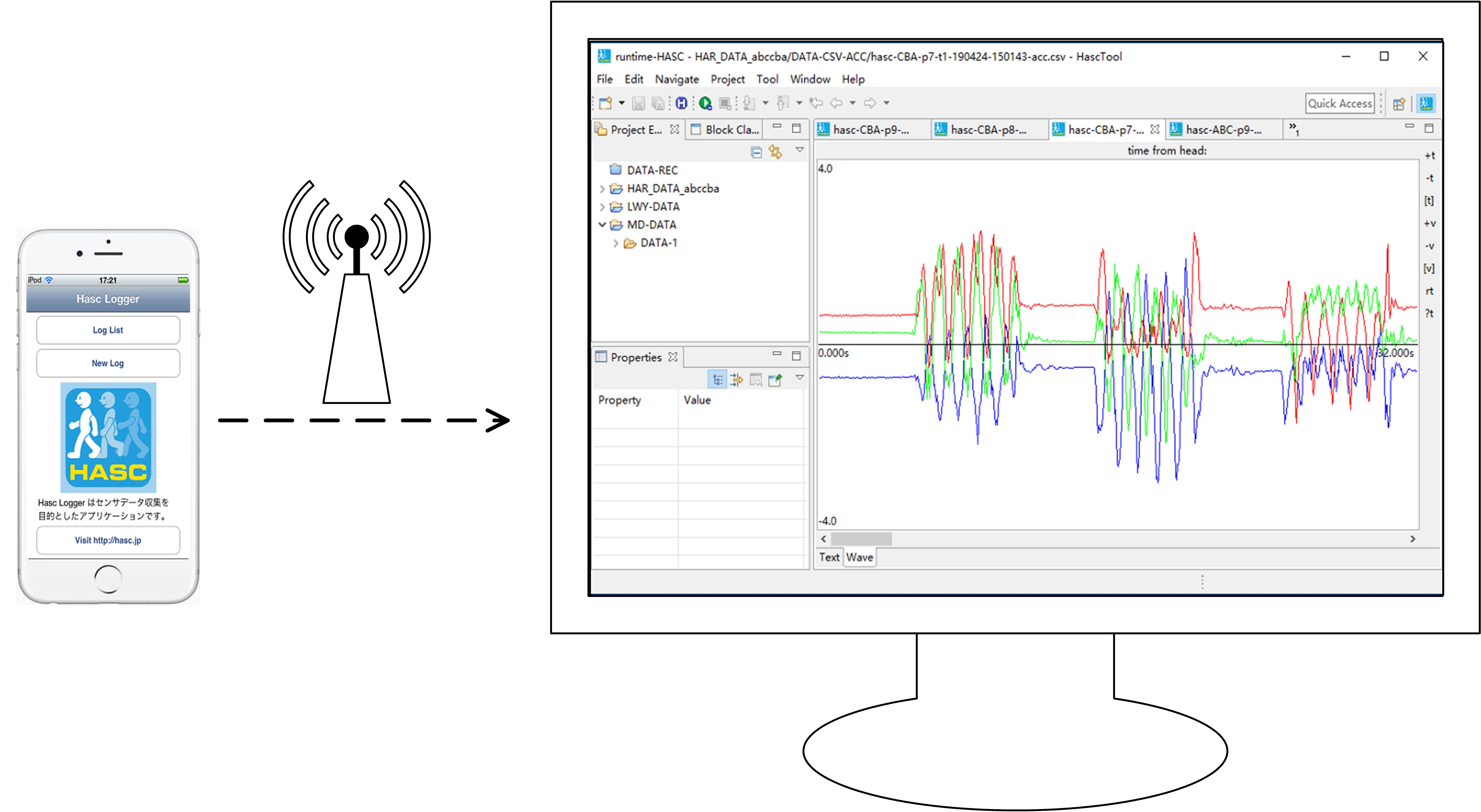}
		\label{Fig. 2(b)}}
	\caption{(a) The smartphone is worn on the right wrist. (b) Uploading the data collected from smartphone to computer terminal by HASC Logger \cite{kawaguchi2011hasc}.}
\end{figure}
\begin{figure}[t]
	\centering
	\includegraphics[width=3.5in]{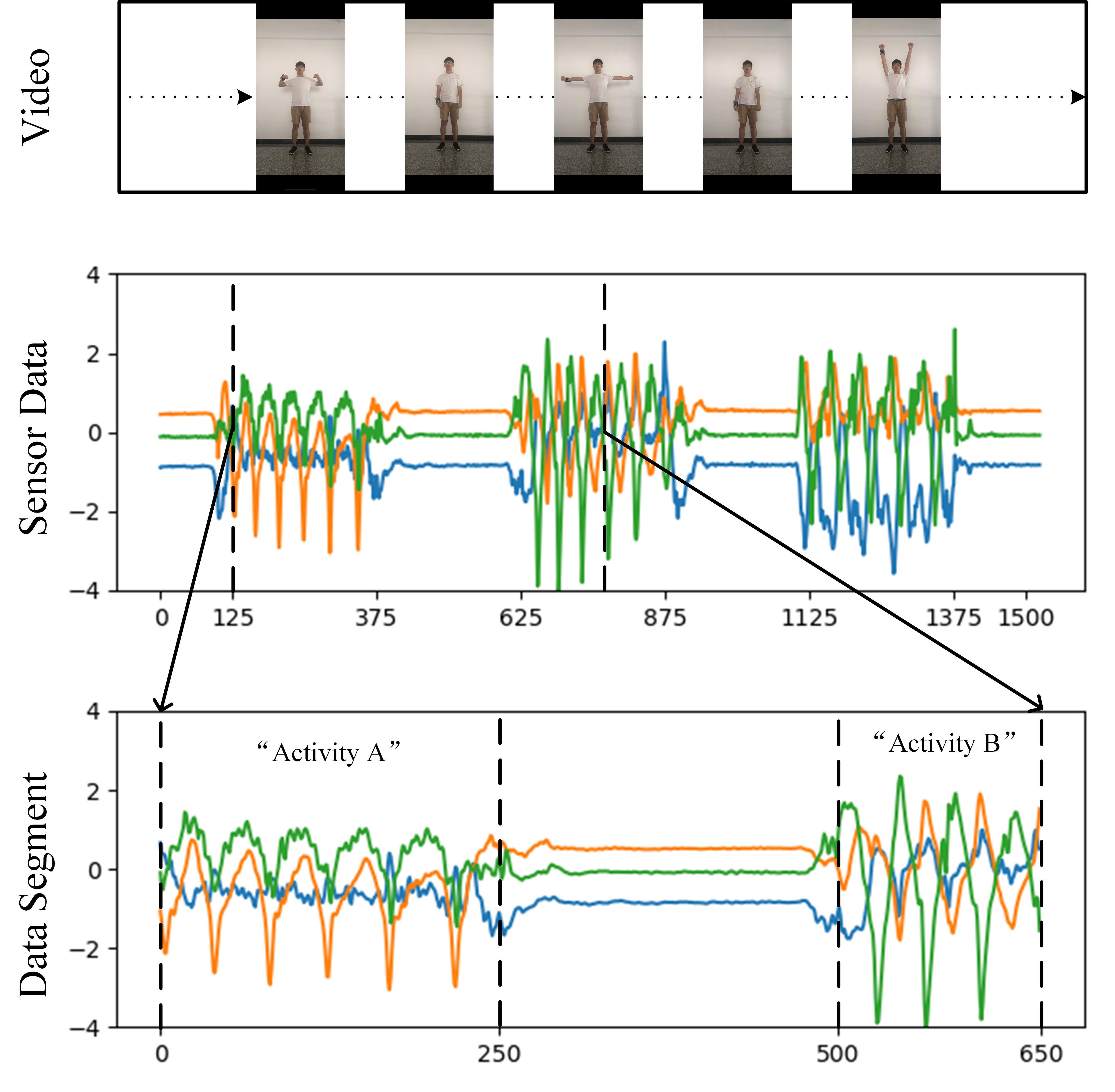}
	\caption{Example of a collected sensor data sequence and a sample labeled "A-B".}
	\label{Fig. 3}
\end{figure}
\subsection{OPPORTUNITY Dataset}
\indent 
\textcolor{black}{To demonstrate the generality and superiority of the proposed method, we try to evaluate the performance on a large public HAR dataset namely OPPORTUNITY, which is devised to benchmark various HAR algorithms. The OPPORTUNITY dataset contains naturalistic human activities collected in a very rich sensor environment where each subject performed daily morning activities. In the paper, we focus on data collected from on-body sensors consisting of inertial measurement units (IMU) and triaxial acceleration. The data, sampled at a frequency of 30Hz, is composed of the recordings of 4 subjects. Each subject was asked to perform 17 different daily activities with 20 repetitions.
\\\indent Without loss of generality, we can use sliding window technique to segment the data into samples with sequential weak labels like the format of the above SWLM data. A large length window of 600 (about 20 seconds) is used to segment the data. As a result, each sample contains one or multiples types of activities. Compared with the SWLM dataset, this processed dataset reflects the concept of the weak label discussed in \cite{wang2019attention}, in which one sample contains not only the target activity but also a large amount of background activities. More than that, one sample contains multiple types of activities that can be regarded as the combination between the target activity and the background activity. For example, we can segment a sequence and obtain a sample that contains the activity “open door” and “close door”. In this case, the activity “close door” can be seen as background activity relative to the target activity “open door” when the model is trying to recognize “open door”. }
\section{Model}
\indent
The aim of the proposed model is to recognize and locate multi-activity types in weakly labeled sensor data. The model consists of CNN, LSTM and attention module. The CNN plays a role of feature extractor that acquires the feature vectors from sensor data. The combination of LSTM and attention module perform twofold functions including determining the location supposed to be paid attention to and generating classification result that matches the corresponding location. Utilizing above methods that mainly inspired by \cite{xu2015show}, we can visualize “where” and “what” the attention focus on. 
\subsection{CNN: Feature Extractor}
\indent
CNN, which has great potential to identify the various salient patterns of HAR’s signals, is used in order to extract features from the raw inputs. CNN maps the input to a set of feature vectors by convolutional kernels: 
\begin{equation}
\mathrm{a}=\left\{a_{1}, a_{2}, \ldots, a_{L}\right\}, a_{i} \in R^{D}
\end{equation}
where $L$ denotes the numbers of feature vectors. Each of vector $a_i$ is D-dimensional representation corresponding to the sensor data.
\\
\indent
A classical CNN architecture consists of convolutional layers, pooling layers and fully connected layer. In our model, the above feature vectors are not fed into the fully connected layer, because we do not need to output the classification probability at this stage. Here the feature vectors are used as the input of attention module and LSTM. The CNN plays a role of an encoder. \textcolor{black}{In the case of the sequential weakly labeled dataset, the shorthand description of the CNN feature extractor is: $Conv_1(16)-Pool_1-Conv_2(32)-Pool_2-Conv_3(64)-Pool_3-Conv_4(128)$, where $Conv_s(k)$ denotes a convolutional layer $s$ with $k$ feature maps (i.e. the number of filters) and $Pool$ a max pooling layer. We apply 1D convolution along temporal dimension and regard the sensor modalities as the channels, which can ensure the following attention mechanism can compute the weighted score along temporal feature vectors. Referring to the setting in the literature \cite{yang2015deep, ordonez2016deep}, we set the kernels size to 5. }
\begin{figure}[t]
	\centering
	\includegraphics[width=2.5in]{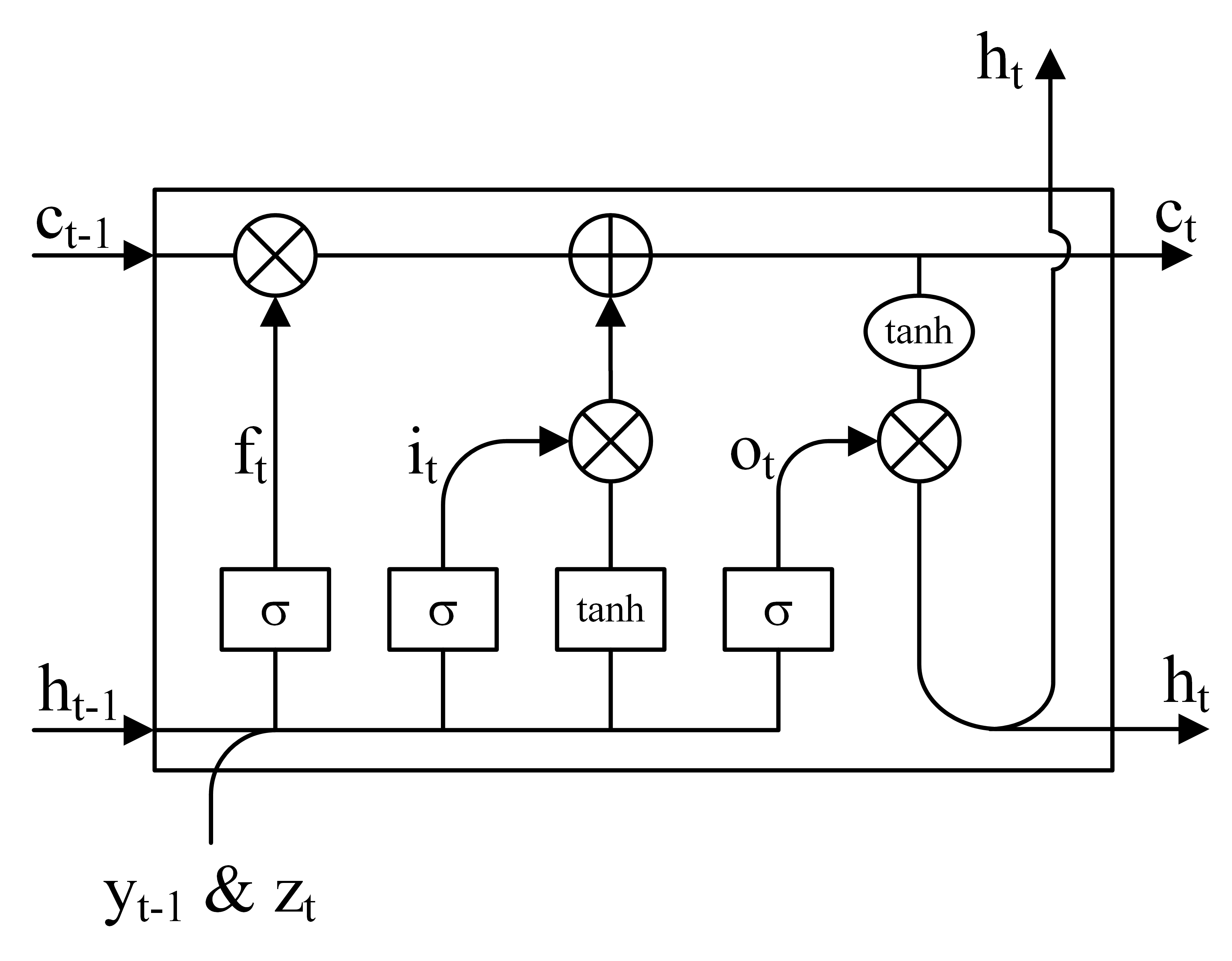}
	\caption{A LSTM cell with input gate, forget gate and output gate.}
	\label{Fig. 4}
\end{figure}
\subsection{LSTM: Decoder}
\indent
LSTM network, introduced by Hochreiter \& Schmidhuber \cite{hochreiter1997long}, is a special type of RNN. LSTM has the form of a chain of repeating modules of neural network, which can be utilized for multi-activity recognition. The implementation of LSTM is presented as (Fig.4): 
\\
\begin{equation}
\left\{\begin{array}{l}{\mathbf{i}_{\mathrm{t}}=\sigma\left(\mathrm{W}_{\mathrm{i}} \mathrm{y}_{\mathrm{t}-1}+\mathrm{U}_{\mathrm{i}} \mathrm{h}_{\mathrm{t}-1}+\mathrm{Z}_{\mathrm{i}} \mathrm{z}_{\mathrm{t}}+b_{i}\right)} \\ {f_{\mathrm{t}}=\sigma\left(\mathrm{W}_{f} \mathrm{y}_{\mathrm{t}-1}+\mathrm{U}_{f} \mathrm{h}_{\mathrm{t}-1}+\mathrm{Z}_{\mathrm{f}} \mathrm{Z}_{\mathrm{t}}+b_{f}\right)} \\ {\mathrm{c}_{\mathrm{t}}=\mathrm{f}_{\mathrm{t}} \mathrm{c}_{\mathrm{t}-1}+\mathrm{i}_{\mathrm{t}} \tanh \left(\mathrm{W}_{\mathrm{c}} \mathrm{y}_{\mathrm{t}-1}+\mathrm{U}_{\mathrm{c}} \mathrm{h}_{\mathrm{t}-1}+\mathrm{Z}_{\mathrm{c}} \mathrm{z}_{\mathrm{t}}+b_{c}\right)} \\ {o_{\mathrm{t}}=\sigma\left(\mathrm{W}_{o} \mathrm{y}_{\mathrm{t}-1}+\mathrm{U}_{o} \mathrm{h}_{\mathrm{t}-1}+\mathrm{Z}_{\mathrm{o}} \mathrm{z}_{\mathrm{t}}+b_{o}\right)} \\ {h_{\mathrm{t}}=o_{\mathrm{t}} \tanh \left(c_{\mathrm{t}}\right)}\end{array}\right.
\end{equation}
where $\mathrm{i}_{\mathrm{t}}, f_{\mathrm{t}}, \mathrm{c}_{\mathrm{t}}, o_{\mathrm{t}}, h_{\mathrm{t}}$ denote the input, forget, memory, output and hidden state of the LSTM respectively. $W, U, Z$ and $b$ are weight matrices and biases learned in the training phase.
\\
\indent
\textcolor{black}{In the previous HAR methods \cite{ordonez2016deep, kumar2018multimodal, kim2018deepgesture}, the LSTM is used to encode the input data or the feature vectors generated from earlier sub-networks, and then feed the encoded vectors to the subsequent classifying networks in order to output classification results. In the paper, the LSTM module plays a role of decoder that generates one classification results at each step $t$ conditioned on the context vector $z_{t}$, the preceding hidden state $h_{t-1}$ and the previously generated classification result $y_{t-1}$.} 
\\\indent The structure of the combination of LSTM and attention module is shown in Fig. 5. The LSTM enables the attention to become “recurrent”, because the hidden state varies as the output RNN advances in its output sequence. That is to say, “where” the network looks next depends on the sequence of classification results that has already been generated. 
Repeating attention on the samples of weakly labeled sensor data containing multiple types of activities can implement the multiple activities recognition task. Besides, it can locate the target activities with sequential weak labels by weighing up the relative importance of different location of sensor data.
At step $t$, the model produces an attention map of the current focused activity sample and its corresponding weighted feature vectors (i.e. the context vector $z_{t}$) by the attention mechanism. The context vector $z_{t}$, which is a dynamic representation of the relevant part of the sensor data input, is computed from the feature vectors $a_{i}$. Then the attention feature vectors replace the original feature vectors produced from the raw sensor data by CNN to participate in the loops of LSTM and then be fed into the classifier. \textcolor{black}{Since the number of time-step should exceed that of the interesting activities contained in the longest sequential weakly labeled sample, it is set to 10 in our implementation. The number of LSTM cells is equal to the size of time-step and the memory units is 128. All cells use logistic sigmoid function as the gate activations, and hyperbolic tangent for the other activations. Each LSTM cell takes as inputs the connection of the previously generated classification result and the context vector, and the hidden state of the previous time. The initial memory state and hidden state of the LSTM are prebuilt by an average of the feature vectors fed through two different multi-layer perceptron:
	\begin{equation}
	c_{0}=f_{i n i t . c}\left(\frac{1}{L} \sum_{i}^{L} a_{i}\right), h_{0}=f_{i n i t . h}\left(\frac{1}{L} \sum_{i}^{L} a_{i}\right)
	\end{equation}
The equation (3) indicates that in the first step (t=1), the weight score is based on the feature vector $a_{0}$  totally.}
\\
\indent
At the end of each time step, a deep output layer (i.e. a dense layer) \cite{pascanu2013construct} $f$ is used to compute the output probability,  saying it is a classifier which cues from the context vector and the hidden state:
\begin{equation}
\mathrm{p}\left(y_{t} | a, y_{t-1}\right) \propto f\left(z_{t}+h_{t}\right)
\end{equation}
\begin{figure}[t]
	\centering
	\includegraphics[width=3in]{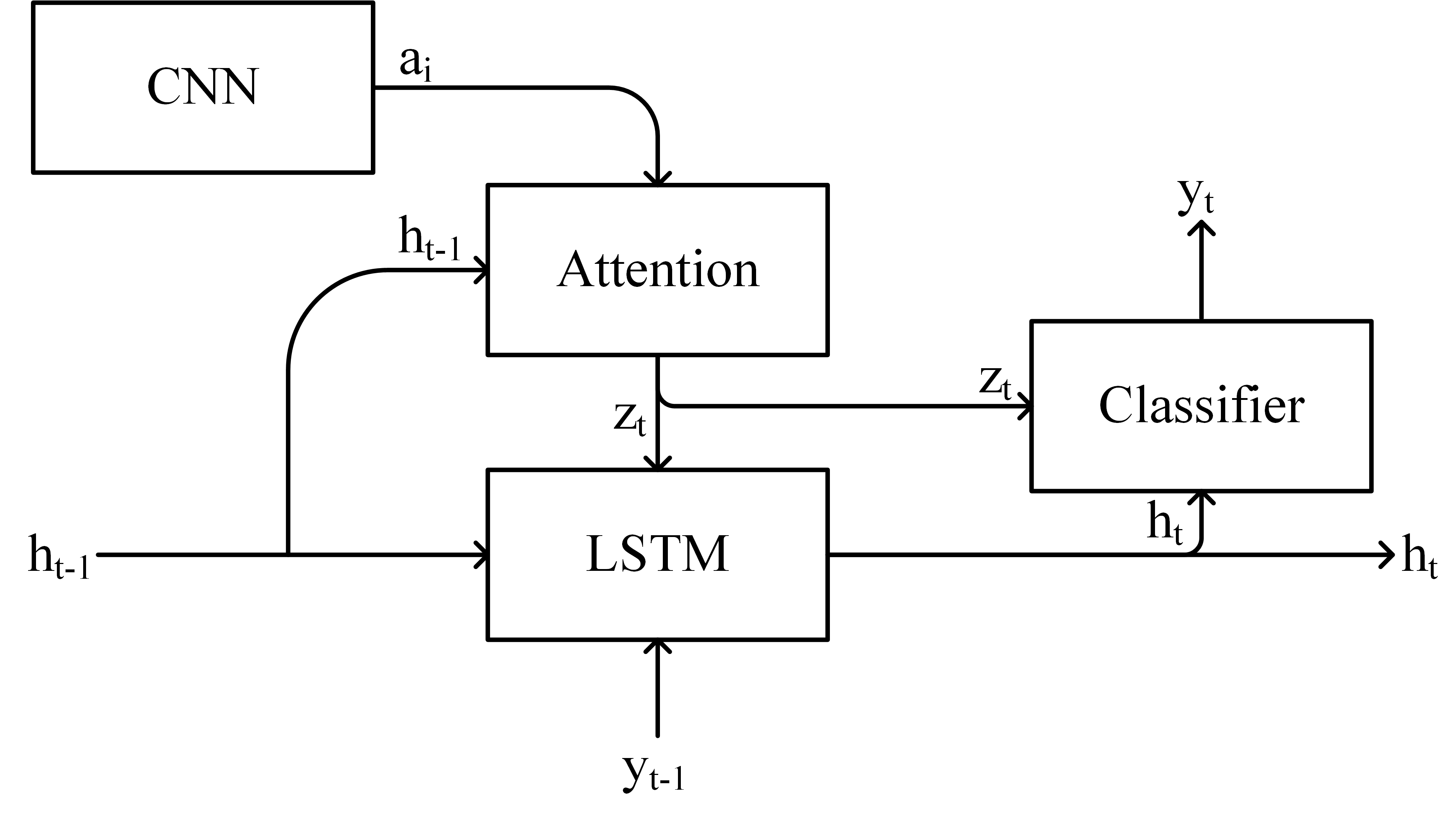}
	\caption{The structure of the combination of LSTM and attention module.}
	\label{Fig. 5}
\end{figure}
\subsection{Attention Module}
In this section, we discuss the details of attention mechanism, that is to say how the context vectors are computed from the feature vectors $a_{i}, i=1,…,L$ corresponding to the features extracted at different temporal locations of sensor data. \textcolor{black}{The attention mechanism produces a positive weight $\alpha_{ti}$ which measures the relative importance of the feature vector $a_i$ for each temporal location $i$. Closely following the one used in \cite{xu2015show}, the weight $\alpha_{t i}$ of each feature vector $a_i$ is computed by using a multi-layer perceptron conditioned on the previous hidden state $h_{t-1}$:}
\begin{equation}
	e_{t i}=W_{a t t}\left(\hat{a}_{i},\left(W_{h} h_{t-1}+b_{h}\right)\right)
\end{equation}
\textcolor{black}{where the $\hat{a}_{i}$ is the projection of the feature vector $a_{i}$ in order to have the same dimension with the hidden state $h_{t-1}$. $W_{att}$, $W_h$ and $b_h$ are the learned weight matrices and biases. The $e_{t i}$ is the weight score of temporal location $i$ and the $t$ is referred to time step.} After the computing process, we have a set of score $S\left(a_{i}, h_{t-1}\right)=\left\{e_{t 1}, e_{t 2}, \ldots, e_{t L}\right\}$,  which are then normalized into $A_{t}=\left\{\alpha_{t 1}, \alpha_{t 2}, \dots, \alpha_{t L}\right\}$ by a softmax function:
\begin{equation}
\alpha_{t i}=\frac{\exp \left(e_{t i}\right)}{\sum_{j=1}^{L} \exp \left(e_{t j}\right)}
\end{equation}
\\ \indent
Then the normalized weights $\alpha_{ti}, i=1,...,L$ are used to produce the context vector $z_{t}$ by element-wise weighted averaging as proposed by \cite{bahdanau2014neural}:
\begin{equation}
z_{t}=\sum_{i=1}^{L} \alpha_{t i} \cdot a_{i}
\end{equation}
\\ \indent
In essence, 
the methods computing the soft attention weighted feature vectors are based on a deterministic attention model, 
which discredits irrelevant activity information by multiplying the corresponding features map with a lower weight. Due to different weighted parameters, the noticeable attention part is enhanced while the less significant attention part is weakened. By this way, the weakly labeled data can be effectively recognized. Besides, the whole model is differentiable, which make an end-to-end training feasible by utilizing standard back-propagation.
\subsection{Optimization}
\indent 
We utilize a doubly stochastic regularization \cite{xu2015show} that encourages the model to pay attention equally to different parts of the sensor data. At time $t$, the attention at every point sums to 1 (i.e. $\sum_{i} \alpha_{t i}=1$), which potentially result in ignoring some parts of the inputs by decoder. In order to alleviate this, we encourage $\Sigma_{t} \alpha_{t i} \approx \tau$ where $\tau \geq \frac{L}{D}$. So the final loss function is defined as:
\begin{equation}
\mathrm{L}_{d}=-\log (\mathrm{p}(\mathrm{y} | \mathrm{a}))+\sum_{i}^{L}\left(\tau-\sum_{t}^{c} \alpha_{t i}\right)^{2}
\end{equation}
where the $\tau$ is fixed to 1. 
\subsection{Localization Method}
\indent The attention mechanism generates the scores by computing the compatibility of the context vectors which contain features extracted by CNN from raw inputs and the hidden states of current step, which indicates the scores should be high if and only if the corresponding parts contain the dominant data category. Taking advantage of this point, one can determine the locations of the target activity in a long sequence of the sensor data.
\\\indent However, the scores generated by the deterministic attention are difficult to be applied in determining locations because the peak of the scores is unstable as discussed in our previous work \cite{wang2019attention}. Thus a localization method is introduced to ameliorate it. As indicated above, we have a set of weighted score  $A_{t}=\left\{\alpha_{t 1}, \alpha_{t 2}, \dots, \alpha_{t L}\right\}$, where $\alpha_{t i}$ is the specific weighted score of the $i-th$ temporal location of the step $t$. A varied width slide window is used to sum up the score within a partial segment:
\begin{equation}
s_{ti}=\left\lbrace \begin{array}{ccl}
\sum\limits_{j=1}^{i+\frac{w}{2}}\alpha_{tj} & \mbox{for} & i<\frac{w}{2}
\\
\sum\limits_{j=i-\frac{w}{2}}^{i+\frac{w}{2}}\alpha_{tj} & \mbox{for} & \frac{w}{2}\leq i\leq n-\frac{w}{2}
\\
\sum\limits_{j=i-\frac{w}{2}}^{n}\alpha_{tj} & \mbox{for} & i>n-\frac{w}{2}
\end{array}\right. 
\end{equation}
where the localization score $s_{ti}$ is corresponding to the summation of the weighted score around the temporal location $i$. The range of this calculation is equal to the slide window width, which is varied as the temporal location i changes. \textcolor{black}{The maximum of window size $w$ is defined as:
\begin{equation}
w = \frac{f}{p^{s}}*d
\end{equation}
where the $f$ is the sampling rate and the $p$ is pooling size. The $s$ is the number of pooling layers and the $d$ is the average duration of the activities. In our implementation of the OPPORTUNITY dataset, the pooling size is 2 and the number of pooling layers is 3. Besides, according to ground truth label, the average duration of the activities is about 3 seconds. Consequently, the maximum of window size $w$ is set to 12.} 
The total temporal location $n$ is equal to the length of the set of weights score $A_{t}$. Then we normalize the $s_{ti}$ into $[0,1]$:
\begin{equation}
\overline{s}_{t i}=\frac{s_{t i}-\min _{i} s_{t i}}{\max _{i} s_{t i}-\min _{i} s_{t i}}
\end{equation}
\\\indent We denote the $\overline{s}_{t i}$ as normalized localization score which presents the importance of the location $i$. Finally, the locations with scores $> 0.7$ (i.e. threshold value, an empirical hyper-parameter) are labeled as potential activity of interest.
\section{Experiments}
\indent
The effectiveness of the proposed model is examined on the three public HAR  datasets: the UniMiB-SHAR dataset, our collected SWLM dataset and the OPPORTUNITY dataset. The former is to validate whether our model has the capacity to implement traditional HAR tasks. The last two are to explore the performance of the proposed model on sequential weakly supervised HAR task. 
\\ \indent
The experiments are performed on a workstation with CPU Intel i7 6850k, 64 GB memory, and a NVIDIA GPU 1080ti with 11GB memory. All algorithm is implemented in Python by using the deep learning framework TensorFlow. In the experiments, the number of epoch is set to 100 and Adam optimization method is used to train our model. The learning rate is set to 0.00025 and the input batch size is 50.
\begin{figure}[t]
	\centering
	\subfloat[“running”]{\includegraphics[width=3.5in]{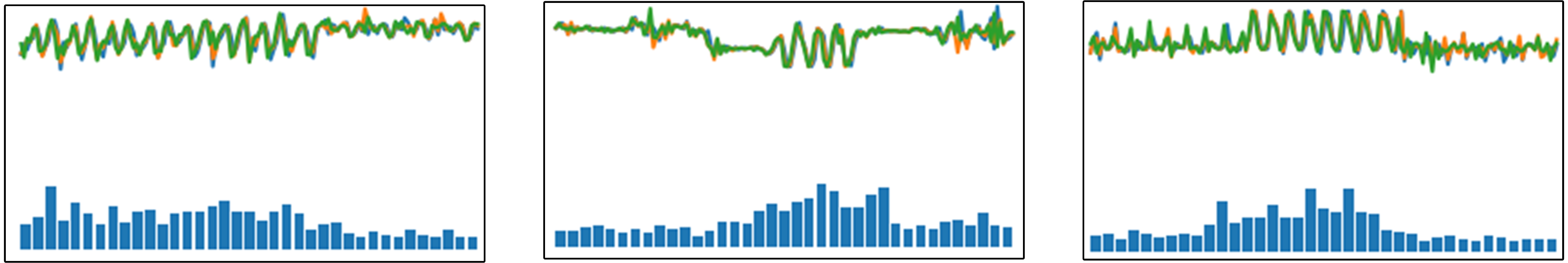}
		\label{Fig. 6(a)}}
	\hfil
	\subfloat[“going down”]{\includegraphics[width=3.5in]{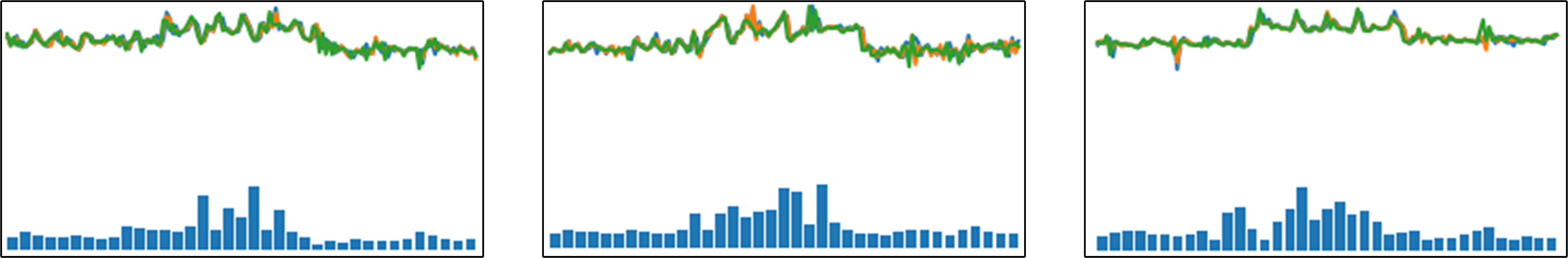}
		\label{Fig. 6(b)}}
	\caption{Some examples of experiments on UniMiB-SHAR dataset.}
	\label{Fig. 6}
\end{figure}
\begin{table}[t]
	\renewcommand{\arraystretch}{1.3}
	\caption{Experiment on UniMiB-SHAR dataset}
	\label{t2}
	\centering
	\begin{tabular}{cc}
		\toprule[1.3pt]
		Model &  Accuracy\\
		\midrule
		CNN & 72.5\% \\
		RAN& 72.8\% \\
		\bottomrule[1.3pt]
	\end{tabular}
\end{table}
\subsection{Experiments on UniMiB-SHAR Dataset}
We compare the experimental results of our model and the baseline CNN model on the UniMiB-SHAR dataset in the metric of classification accuracy. In this experiment, the baseline CNN consists of three convolutional layers with 32, 64, 128 kernels and two max pooling layers between these three convolutional layers. One fully connected layers with 100 units is used, and then a softmax layer is used to output the classification results. The baseline CNN without the fully connected layer is corresponding to the features extractor of our model. 
\\ \indent
The results are shown in Table \uppercase\expandafter{\romannumeral1}.  \textcolor{black}{Our model and the baseline CNN perform comparably well with regard to traditional task that belongs to supervised learning. Our RAN model replaces the final fully connected layer of the fundamental CNN architecture with the attention module. As shown in Fig. 6(a), the attention module can identify the salient activity data areas and enhance their influence, meanwhile suppressing the irrelevant and potentially confusing information in other activity data areas. However, compared with the attention mechanism proposed in \cite{wang2019attention, jetley2018learn}, the soft attention only adds the weights to the features extracted at the end of the CNN pipeline, which potentially weakens the specific features selection capability of attention mechanism. Thus, for this dataset, our model is not superior to the traditional methods, but still performs satisfactorily. Fig. 6 demonstrates the effect of attention mechanism on sensor data, which indicates that the attention scores correspond to the importance of the features extracted from different part of sensor data. Besides, it can be seen that the attention mechanism can selectively focus on the identical features of the same type of activity.}

\subsection{Experiments on SWLM Dataset}
In this experiment, we use a CNN introduced in Section IV as feature extractor. A fully connected layer and a softmax layer are added to build the baseline CNN. The five different experiments are designed according to the types of activities which appears in training set and test set, as shown in Table  \uppercase\expandafter{\romannumeral2}. Throughout above experiments, we can validate the effectiveness of the RAN model in HAR tasks, and explore its potential applications for activity localization tasks.  
\\\indent\textit{Case 1:} In this case, the training set and test set only contain the sensor data samples consisting of one weakly labeled activity, which can been seen as a traditional recognition task. Due to the simple constitution of the activities, it is easy for our RAN model and the baseline CNN to extract distinct features from the dataset. So both methods can achieve an almost 100\% accuracy on this traditional task. Moreover, the attention mechanism of our model exerted on the dataset is shown in Fig. 7. Note that the purpose of our work is not to pursue higher performance in recognizing activity, but to develop a sequential activity recognition and localization method which can detect and locate accurately each activity in one sample, as mentioned in the following experiments.  
\begin{figure}[t]
	\centering
	\includegraphics[width=3.5in]{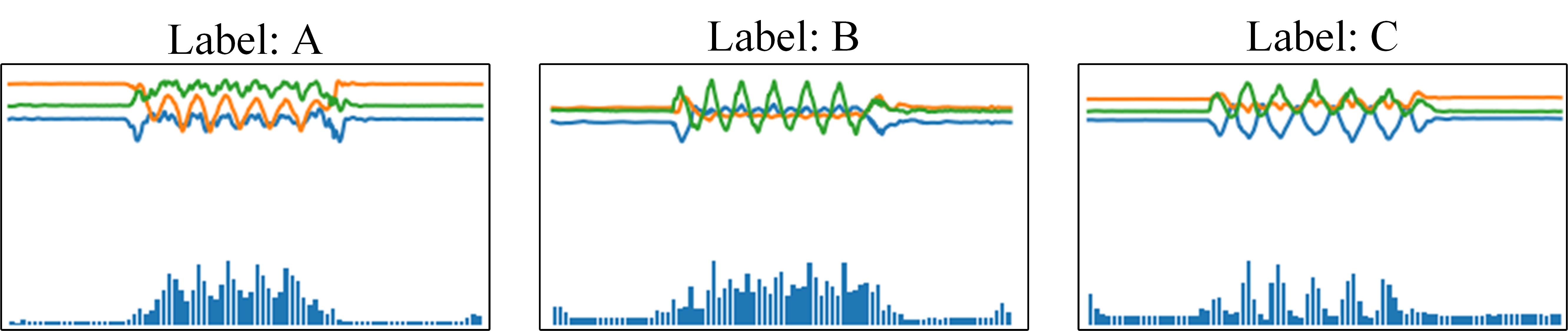}
	\caption{Examples of experiment case 1. The attention mechanism of our model weights the features of different locations.}
	\label{Fig. 7}
\end{figure}  
\begin{figure}[t]
	\centering
	\subfloat[]{\includegraphics[width=3in]{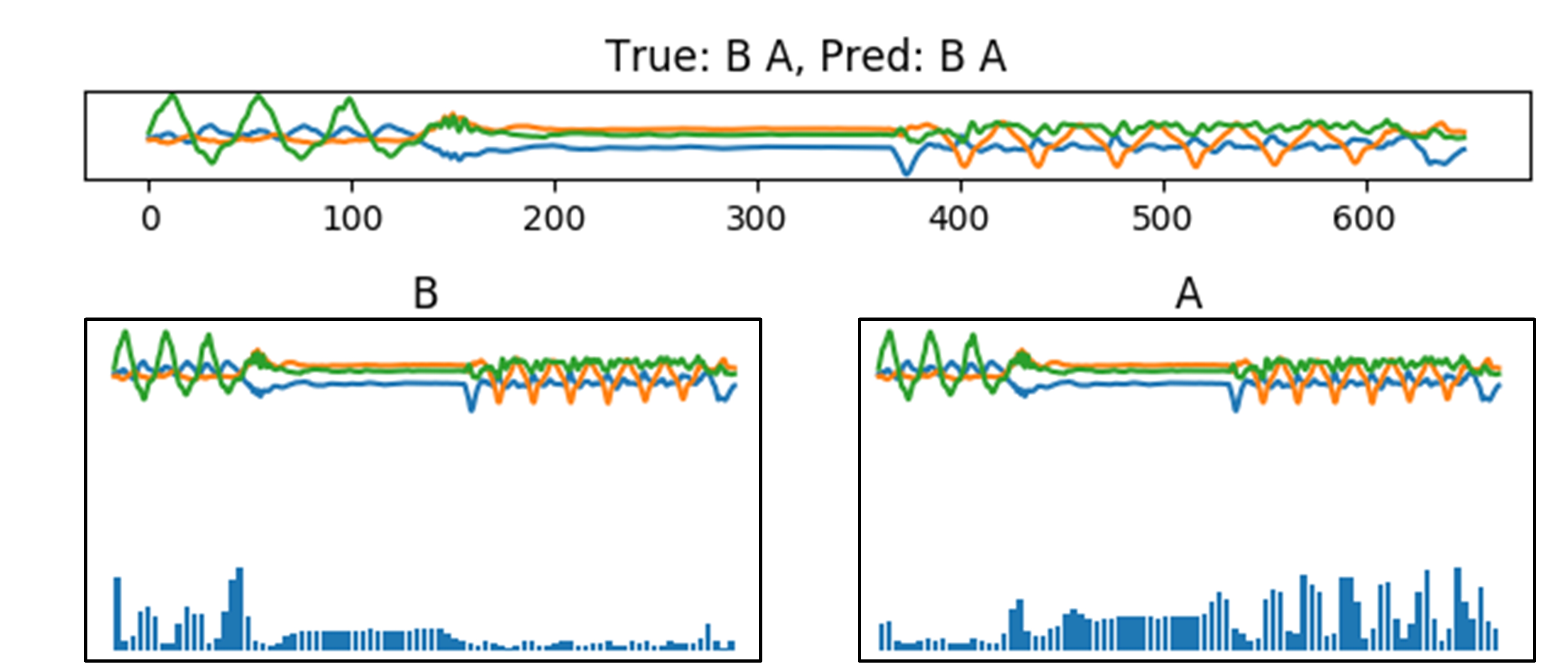}
		\label{Fig. 8(a)}}\hfil
	\subfloat[]{\includegraphics[width=3in]{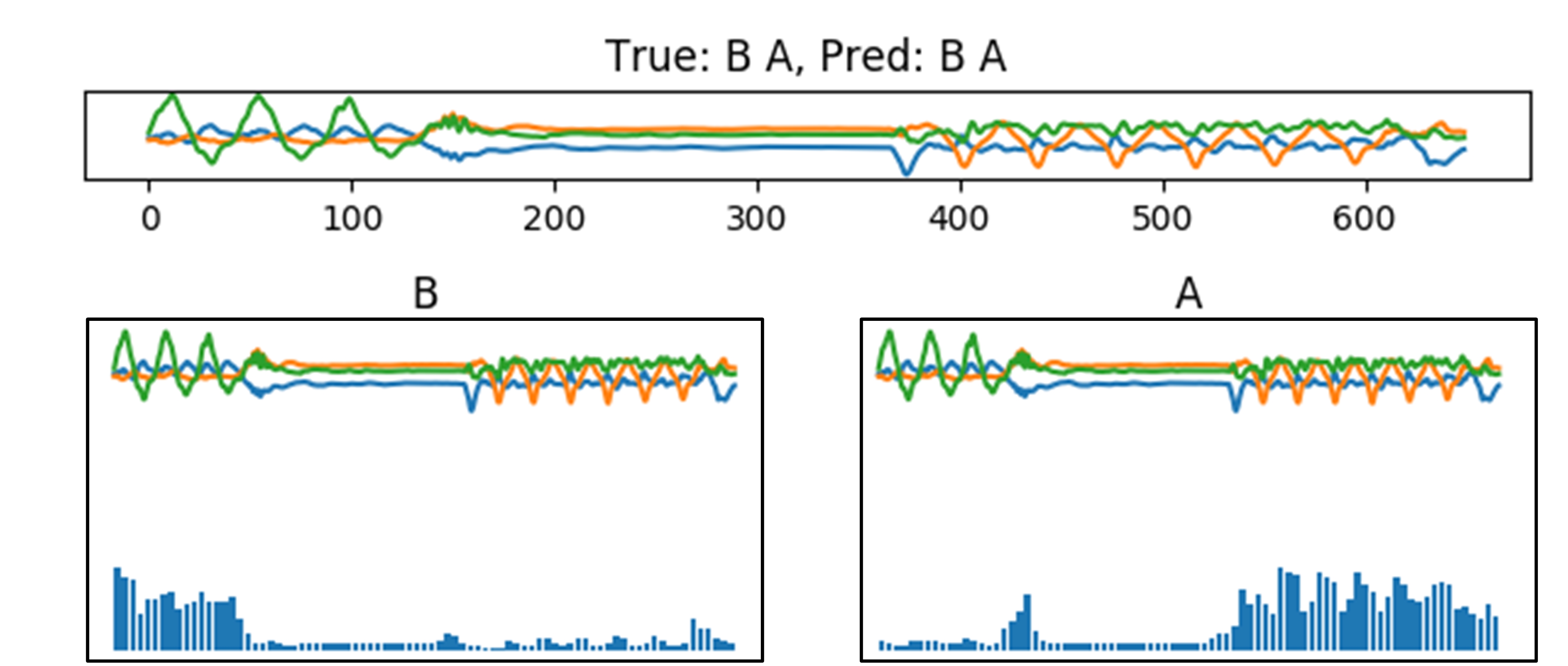}
		\label{Fig. 8(b)}}
	\caption{(a) Examples of experiment case 2. At each step, the model produces an attention map revealing where the current step is focusing on and an classification result conditioned on the focused part. (b) Examples of experiment case 3. The attention maps become more distinct.}
\end{figure}
\\\indent\textit{Case 2:} This case is to test whether our model can detect activities in situation that the samples of training set contain multiple types of labels. \textcolor{black}{Notably, since the baseline CNN model is unable to respectively extract features of every activity with different labels simultaneously, the baseline CNN does not directly perform the sequential recognition task for the samples annotated with sequential weak labels. We can compel the baseline CNN to implement the recognition task by annotating the multi-activity samples as a new label (e.g. marking “A-B” as “D”, “B-A” as “E”, etc.), and as a result it degenerates into a traditional supervised learning task, whose purpose is to classify six types of activities. Using this idea, the baseline CNN can obtain a 99.2\% classification accuracy, but the annotation processes become much harder (thus we mark the results with asterisk). Actually, the baseline CNN does not recognize every activity contained in one segment, but classifies simply this sample as a whole. On the contrary, the RAN model at one time only use the features of an activity in the sequence to perform recognition. Therefore, the RAN model sometimes shows a worse accuracy compared to the baseline CNN. However, the advantage of our model lies in that it can locate every labeled activity in one sample, and in the meantime get a satisfactory classification result.}
\begin{table}[t]
	\renewcommand{\arraystretch}{1.5}
	\caption{Experiments Cases and Results}
	\label{t3}
	\centering
	\begin{tabular}{lllcc}
		\hline\hline
		\multicolumn{1}{c}{} & \multicolumn{2}{c}{Distribution}                                                                                                                              & \multicolumn{2}{c}{Accuracy}                            \\ \cline{2-5} 
		Case                 & \multicolumn{1}{c}{Train}                                                         & \multicolumn{1}{c}{Test}                                                  & \multicolumn{1}{c}{CNN} & \multicolumn{1}{c}{RAN} \\ \hline
		1                    & A, B, C                                                                           & A, B, C                                                                   &              100\%           &               100\%                \\ \hline
		2                    & \begin{tabular}[c]{@{}l@{}}A-B, B-A, C-A,\\ A-C, B-C, C-B\end{tabular}            & \begin{tabular}[c]{@{}l@{}}A-B, B-A, C-A,\\ A-C, B-C, C-B\end{tabular}    &             *99.2\%            & 99.0\%                       \\ \hline
		3                    & \begin{tabular}[c]{@{}l@{}}A, B, C,\\ A-B, B-A, C-A,\\ A-C, B-C, C-B\end{tabular} & \begin{tabular}[c]{@{}l@{}}A, B, C,\\ A-B, B-A, C-A,\\ A-C, B-C, C-B\end{tabular} &           *98.9\%              & 98.5\%                       \\ \hline	
		4                    & \begin{tabular}[c]{@{}l@{}}A, B, C \\ A-B, A-C, B-C \end{tabular}                                                                    & B-A, C-A, C-B                                                             &           -              & *85.6\%                        \\\hline  
		5                   & \begin{tabular}[c]{@{}l@{}}A, B, C, A-B\\  B-A, A-C, C-A\end{tabular}                     & B-C, C-B                                                                  &           -              & -                             \\ \hline\hline
	\end{tabular}
\end{table}
\begin{figure*}[t]
	\centering
	\includegraphics[width=7in]{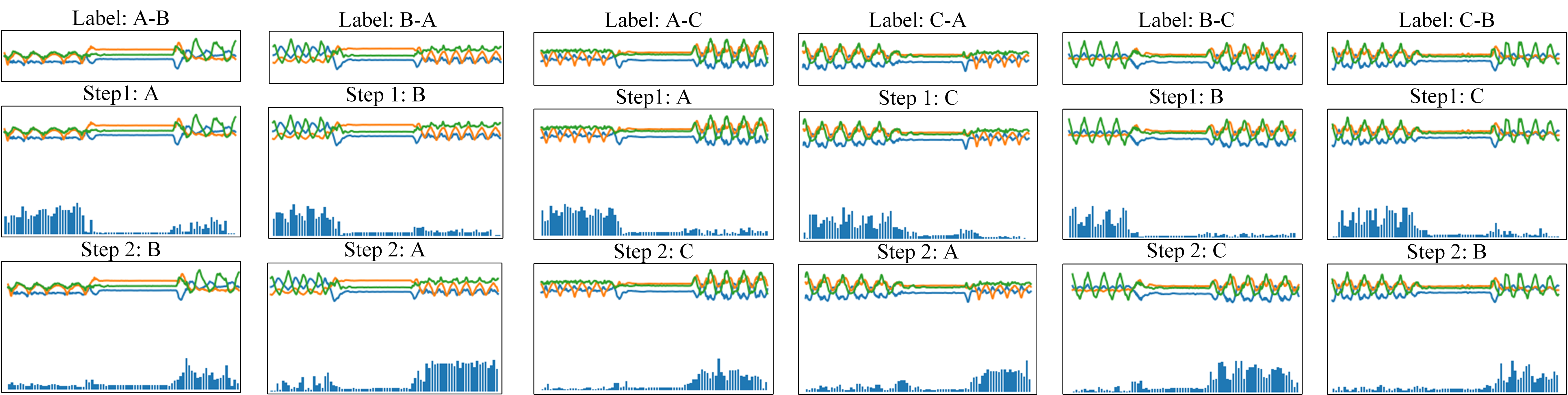}
	\caption{Some additional examples of experiment case 3. The attention module has better weighting capability and the generated attention maps become more clear, due to the addition of single
		activity. }
	\label{Fig. 9}
\end{figure*}
\begin{figure}[t]
	\centering
	\includegraphics[width=3in]{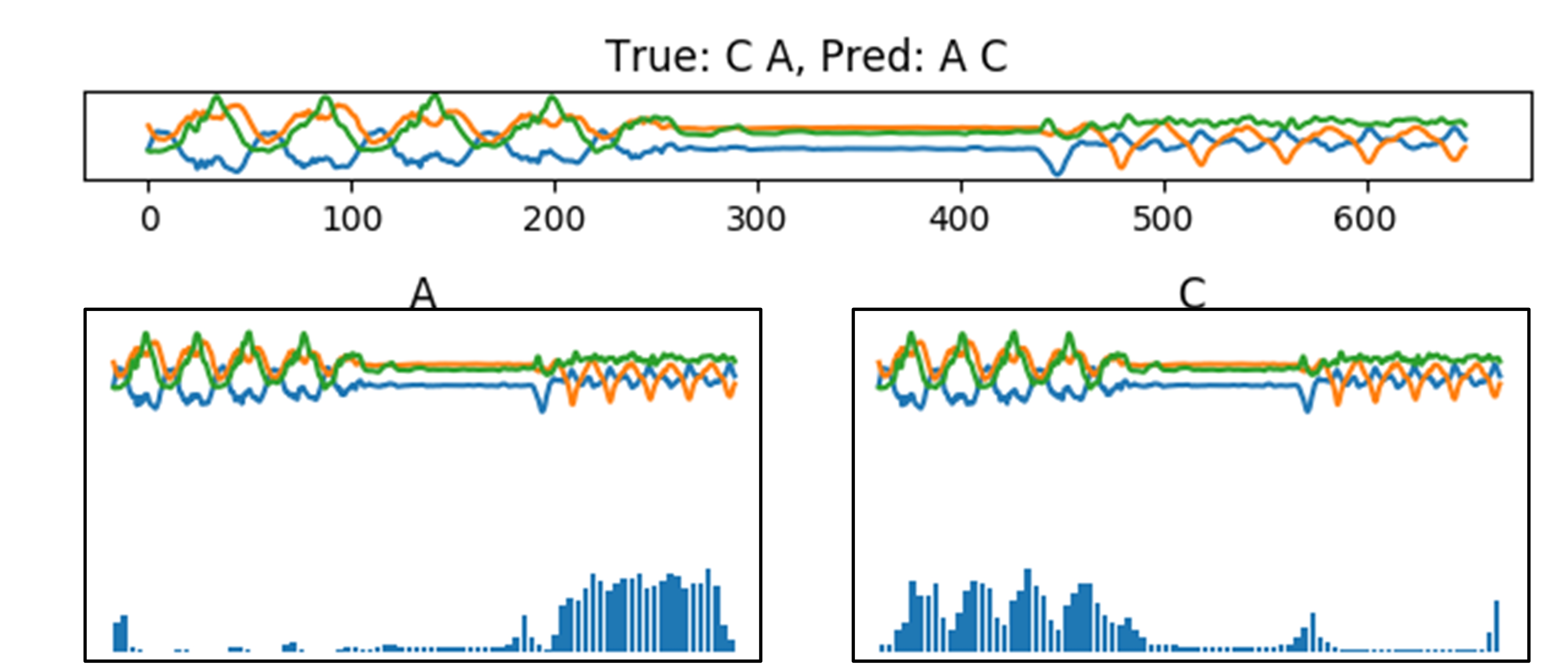}
	\caption{A example of experiment case 4: The model can recognize the activities with correct classification results but reverse order.}
	\label{Fig. 10}
\end{figure} 
\\ \indent Fig. 8(a) shows the recognition examples of our RAN model. At each step, the model produces an attention map revealing where the current attention is focusing on and a classification result corresponding to the focused activity simultaneously. In spite of almost the same classification accuracy, our model is very different from the baseline CNN, which treats these multi-activity samples as one whole. Our RAN model can recognize every labeled activity in one sample, and in the meantime get a satisfactory classification result. We stress that our RAN model can be used to handle the weakly labeled sequential activity dataset, which differs completely from the traditional supervised learning techniques requiring the accurate bounding boxes for annotating the training dataset. Thus, the complex and laborious process of manual annotation work can be greatly alleviated by our  model, which can automatically crop the desired activity by utilizing the recurrent attention mechanism to impose the  proper weights on the sequential activity data. That is to say, we can determine the accurate location of every labeled activity by visualizing the attention maps. However, as shown in Fig. 8(a), the attention maps can not track the location of the desired activity very well, which will be solved in the next case.
\\
\indent \textit{Case 3:} For the case 2, the attention maps can not match accurately the location of every activity in one sample. As the samples in the dataset only contain the multiple labeled activities, the attention module can not learn the features of single activity. In this case, we combined the data samples of single activity (i.e. “A”, “B” and  “C”) and multi-labeled activity (i.e. “A-B”, “B-A”, “A-C ”, “C-A”, “B-C” and “C-B”) into the dataset, in order to further explore the effectiveness of our model. Table \uppercase\expandafter{\romannumeral2} indicates that the baseline CNN and our RAN model still can obtain  98.9\% and 98.5\% classification accuracy respectively, which are almost the same. Fig. 8(b) shows that compared to the case 2, the attention module has better weighting capability and the generated attention maps become more clear, due to the addition of single activity. More clear attention maps facilitates the determination of activity location, which makes it possible to automatically crop the regions of interest for acquiring the labeled HAR dataset by roughly marked the sequential activity data. The location part will be discussed in section D.
\\ \indent \textit{Case 4:} We further try to test whether our model can recognize the sequential activity which does not exist in the training set. In this case, a relatively simple recognition task is proposed. We perform the recognition for sequential activity of reverse order, saying that the training set contains “A”, “B”,  “C”, “A-B”, “A-C” and “B-C” while the test set contains “B-A”, “C-A” and “C-B”. As Fig. 10 shows, the attention can focus on the right location of the current activity at each step, but the LSTM submodule, in charge of generating captions for sequential activity, outputs sequential classification results with reverse order. After labeling reversely the sequential activities of classification results, we still can obtain a classification accuracy of 85.6\%. The results indicate that our model can recognize the sequential activity with reverse order, which never exists in the training set.
\\ \indent \textit{Case 5:} In this case, the dataset is reorganized as follows: the training set contains “A”, “B”, “C”, “A-B”, “B-A”, “A-C” and “C-A”, and the test set contains “B-C” and “C-B”. We continue to explore whether our model can recognize the sequential activity which never appears in the training set. The case 4 can be seen as a special case, where the reverse order condition holds. The result indicates that the data samples of "B-C" and "C-B" can not be recognized and our RAN model fails to generate new captions for sequential classification results. Actually, the LSTM submodule can not remember the sequential activity which never appears at the training stage, and the attention module does not learn how to impose the weights on these samples. That is to say, to realize accurate annotation via the RAN model, the desired sequential activities have to be roughly segmented and trained in advance.
\\\indent On the whole, the above cases indicate that, \textcolor{black}{unlike the baseline CNN which can only handle traditional supervised leaning task, our RAN model is able to recognize and locate every activity contained in a long sequence and achieves an almost the same classification accuracy with the CNN. In other words, attention can tell where to focus along a long sequence of sensor data. Our RAN model can aid the annotator to perform annotation via skimming through raw sensor data and effectively label all activity instances. The only possible obstacle to our model is to recognize the sequential weakly labeled data samples which never appears at the training stage. This is one common problem for supervised learning techniques, which can be easily solved by roughly segmenting and training the desired sequential activity in advance for the RAN model. In addition, our model can recognize the data samples with reverse order label, which indicates the limitation of LSTM does not impede the implementation of the attention.}
\subsection{Experiments on OPPORTUNITY Dataset}
\indent\textcolor{black}{We perform two experiments on the OPPORTUNITY dataset to validate the effectiveness of our RAN model. Firstly, we compare CNN, DeepConvLSTM [15] and our RAN model in the metrics of classification accuracy. Actually, CNN and DeepConvLSTM cannot recognize or locate  multi-activity types in sequential sensor data segment.
For comparison, we have to use a shorter length window of 64 to segment the data in order to build a normal HAR dataset with strict labels. That is to say, we compare the three models on strictly labeled HAR dataset. Secondly, a large length window of 600 is used to segment the data, which yields a sequential weakly labeled dataset. The design of CNN is the same to the above SWLM experiment in which the structure of CNN is equal to the feature extractor of our RAN model. The DeepConvLSTM model is rebuilt according to the original paper \cite{ordonez2016deep}. The results shown in Table \uppercase\expandafter{\romannumeral3} indicates that, in strictly labeled case, our RAN model cannot outperform the start-of-art DeepConvLSTM model, due to its simplicity of feature extractor CNN. However, our RAN model is superior to the standard CNN. In addition, this method is very applicable for recognizing and locating sequential weakly labeled data. Although we may compel the traditional methods such as CNN and DeepConvLSTM to implement recognition task by annotating the multi-activity data samples as a new label like the last experiment did, it will greatly increase the burden of sensor data annotation, not to mention the numerous combinations of activities in the OPPORTUNITY dataset. Besides, the experiment result of sequential weakly labeled data recognition on SWLM dataset is superior to that of OPPORTUNITY dataset, because the background activity of the latter is more confused and the adjacent activities contained in some samples are too close. Nevertheless, the experiment result demonstrates that our RAN can obtain 77.5\% classification accuracy on sequential weakly labeled data, which reaches the level that the CNN performing traditional recognition task.}
\begin{table}[t]
	\renewcommand{\arraystretch}{1.5}
	\caption{Experiment on OPPORTUNITY Dataset}
	\label{t4}
	\centering
	\begin{tabular}{llll}
		\toprule[1.5pt]
		Model &  CNN & \cite{ordonez2016deep} & RAN\\
		\midrule
		\textbf{-Strictly Labeled Data-}& & &\\
		Accuracy & 77.8\%& 82.6\%& 78.6\%\\
		\midrule
		\textbf{-Sequential Weakly Labeled Data-}& & & \\
		Accuracy & -& -& 77.5\%\\
		\bottomrule[1.2pt]
	\end{tabular}
\end{table}
\begin{figure}[t]
	\centering
	\centering
	\subfloat[]{\includegraphics[width=3in]{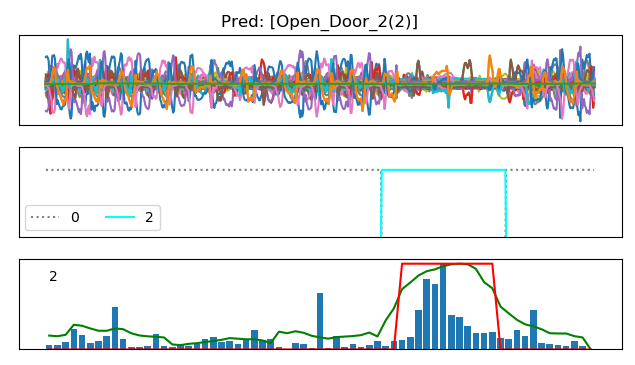}
		\label{Fig. 11(a)}}\hfil
	\subfloat[]{\includegraphics[width=3in]{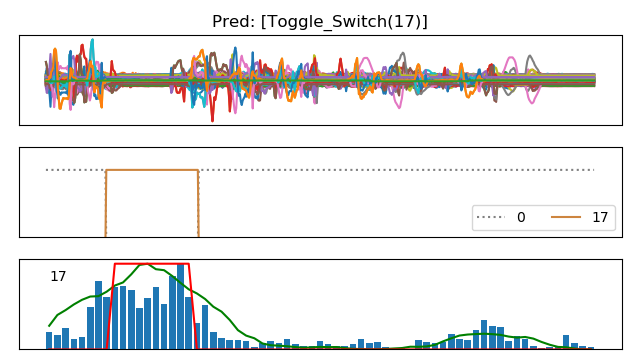}
		\label{Fig. 11(b)}}\hfil
	\subfloat[]{\includegraphics[width=3in]{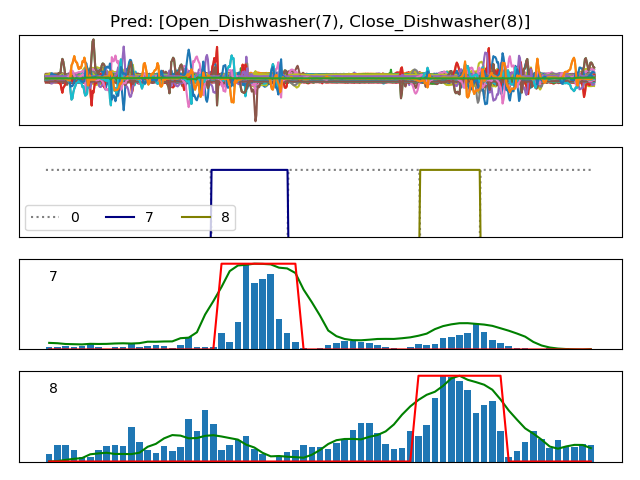}
		\label{Fig. 11(c)}}
	\caption{\textcolor{black}{Visualizing some samples of localization experiments on the OPPORTUNITY dataset. Each sub-figure contains the sensor data diagram, the ground truth and the localization results from top to bottom. The tag "0" is referred to background activity.}}
\end{figure}
\subsection{Localization Experiments}
By utilizing the weighted scores produced by the RAN model, we are able to locate the sequential weakly labeled activity, as shown in the above experimental figures. What's more, we convert the weighted score to the normalized localization score on sequential weakly labeled sample of OPPORTUNITY dataset. \textcolor{black}{As shown in Fig. 11, each sub-figure contains the sensor data diagram, the ground truth and the localization results from top to bottom. The figures show that compared with the weighted score plotted in blue, the green curve of the normalized localization score is more beneficial to locate the labeled activities, because the green curve concentrates on the peak point where the target activity happens more intensively. The red curve is used to mark the partial segment where the normalized localization scores are above the threshold value. The red crops can be regard as the regions of interest, which are close to the scale of ground truth. In addition, it can be seen that the RAN model can not only locate the one activity of interest in weakly labeled samples as shown in Fig. 11 (a) and (b), but also deal with the sequential weakly labeled sample whose segment contains multiple activities as shown in Fig. 11(c).}
\section{Conclusion}     
\textcolor{black}{
One challenge for deep HAR recognition is the collection of annotated or “ground truth labeled” training data. Ground truth annotation is an expensive and tedious
task, in which the annotator has to perform annotation via skimming through raw sensor data and manually label all activity instances. However, the time series data recorded
from multimodal embedded sensors such as accelerometer or gyroscope is far more difficult to interpret than data from other sensors modalities, such as cameras. It requires laborious human efforts to accurately segment and label a target activity from a long sequence of time series sensor data. Therefore, it deserves further research whether we can directly recognize and locate one or multiple target activity from coarsely labeled sensor data. That is to say, our main research motivation is to simultaneously infer multi-activity types from the coarse-grained sequential weak labels and determine specific locations of every target activity with only knowledge of which types of activities contained in the long sequence. It will greatly reduce the burden of manual labeling. 
\\ \indent Hence, we develop a RAN model for multi-activity recognition that can repeatedly pay attention to the activity of interest at each step. The experiments show that our RAN model can perform HAR tasks on sequential weakly labeled data that the traditional deep learning methods can hardly handle. Besides, the RAN model provides a way to locate the activities of interest on the sensors data. Our research also provide a feasible way to apply the attention mechanism to aid to the location and annotation of motion sensor data. Actually, the location of motion sensor data has seldom been researched. Attention mechanism demonstrates its superiority in recognition and location task for motion sensor data, which deserves deeper studies. For example, building a more efficient feature extractor sub-networks for the RAN model and attention mechanism. We put these as our future works.}

\bibliographystyle{IEEEtran}
\bibliography{ref}

\end{document}